\documentclass[aps,twocolumn,showpacs,pra,preprintnumbers,amsmath,amssymb]{revtex4}
\newcommand{\lsim} 
 {\ \raise.35ex\hbox{$<$}\kern-0.75em\lower.5ex\hbox{$\sim$}\ }
\newcommand{\gsim}
 {\ \raise.35ex\hbox{$>$}\kern-0.75em\lower.5ex\hbox{$\sim$}\ }
\newcommand{\bras}[1]{\langle#1|}
\newcommand{\kets}[1]{|#1\rangle}

\newcommand{\mean}[1]{\left<#1\right>}
\newcommand{\means}[1]{\langle#1\rangle}

\def\journal #1#2#3#4{#1 {\bf #2}, #3 (#4)}

\def\PRB{Phys.\ Rev.\ B}
\def\PRL{Phys.\ Rev.\ Lett.}

\def\JPSJ{J.\ Phys.\ Soc.\ Jpn.}

\def\PTP{Prog.\ Theor.\ Phys.}

\usepackage{graphicx}
\usepackage{dcolumn}
\usepackage{bm}
\usepackage{amsmath}
\usepackage{times}
\usepackage[usenames]{color}

\usepackage{ulem}

\begin{document}
\title{
Thermal fractionalization of quantum spins in a Kitaev model: \\
$T$-linear specific heat and coherent transport of Majorana fermions
}
\author{Joji Nasu,$^{1}$ Masafumi Udagawa,$^{2}$ and Yukitoshi Motome$^{2}$} 
 \affiliation{$^{1}$Department of Physics, Tokyo Institute of Technology, Ookayama, 2-12-1, Meguro, Tokyo 152-8551, Japan,\\
$^{2}$Department of Applied Physics, University of Tokyo, Hongo, 7-3-1, Bunkyo, Tokyo 113-8656, Japan}
\date{\today}
\begin{abstract}
Finite-temperature ($T$) properties of a Kitaev model defined on a honeycomb lattice are investigated by a quantum Monte Carlo simulation, from the viewpoint of fractionalization of quantum $S=1/2$ spins into two types of Majorana fermions, itinerant and localized.
In this system, the entropy is released successively at two well-separated $T$ scales, as a clear indication of the thermal fractionalization.
We show that the high-$T$ crossover, which is driven by itinerant Majorana fermions, is closely related with the development of nearest-neighbor spin correlations.
On the other hand, the low-$T$ crossover originates in thermal fluctuations of fluxes composed of localized Majorana fermions, by which the spectrum of itinerant Majorana fermions is significantly disturbed. 
 As a consequence, in the intermediate-$T$ range between the two crossovers, the system exhibits $T$-linear behavior in the specific heat and coherent transport of Majorana fermions, which are unexpected for the Dirac semimetallic spectrum in the low-$T$ limit.
We also show that the flux fluctuations tend to open an energy gap in the Majorana spectrum near the gapless-gapped phase boundary.
Our results indicate that the fractionalization is experimentally observable in the specific heat, spin correlations, and transport properties.
\end{abstract}

\pacs{75.10.Kt, 75.70.Tj, 75.10.Jm}


\maketitle



%
%

%




\section{Introduction}

The fractionalization of electrons in solids is one of the central topics in modern condensed matter physics.
A prototypical example is found in one-dimensional strongly correlated electron systems: charge and spin degrees of freedom in an electron behave as independent particles, which are termed holon and spinon, respectively~\cite{Tomonaga1950}. 
A different form of fractionalization is also anticipated in insulating magnets with geometrical frustration.
For instance, the existence of the elementary excitation carrying a half of spin, named spinon, is predicted in a quantum spin liquid (QSL)~\cite{Wen1991,Nussinov2007,Normand2014}, and emergence of magnetic monopoles is suggested in spin ice systems~\cite{Castelnovo2008}.
Another fractionalization was pointed out in heavy fermion systems as well.
The half residual entropy in the two-channel impurity Kondo system is understood from the fractionalization of $S=1/2$ impurity spin into two Majorana fermions~\cite{Emery1992}.

A quantum spin model, called the Kitaev model, has recently attracted considerable attention in broad areas of research, not only condensed matter physics but also statistical physics and quantum information~\cite{Kitaev06}. 
This model is composed of $S=1/2$ spins with bond-dependent interactions on a honeycomb lattice.
Such peculiar interactions were suggested to be realized in the systems with strong spin-orbit coupling, such as iridium oxides~\cite{Jackeli2009}.
The most striking feature of this model is that it is exactly solvable due to the existence of $Z_2$ conserved quantity on each hexagon, termed flux. 
The ground state dictates both gapless and gapped QSL phases depending on the exchange coupling constants.
The exact solution is provided by representing $S=1/2$ spins by two types of Majorana fermions: one is localized and composes the fluxes, and the other forms itinerant bands~\cite{Chen2007,Feng2007,Chen2008}. The latter itinerant Majorana fermions determine the excitation spectrum in the QSLs.
Thus, the fractionalization of spins into Majorana fermions is not just a mathematical tool but physically important in the Kitaev model.

A natural question arising here is how high-temperature ($T$) paramagnetic spins are fractionalized into Majorana fermions when cooling the system. 
Considering the fact that the spin-charge separation in the one-dimensional electron systems plays a key role in comparison with experiments, it is crucial to elucidate the thermal fractionalization for experimental exploration of the QSL physics.
The thermodynamic properties in the Kitaev model and its extensions have been studied, mainly for explaining the magnetism in iridium oxides~\cite{Chaloupka2010,Reuther2011,Chaloupka2013,Nasu2014,Nasu2014a,Nasu2014b}, 
but the signature of fractionalization at finite $T$ was not addressed in most of the previous studies.
Among them, however, the authors pointed out the significance of fractionalization in a peculiar phase transition at finite $T$ in a three-dimensional extension of the Kitaev model~\cite{Nasu2014,Nasu2014a}: the phase transition is governed by thermal excitations of localized Majorana fermions.
Nevertheless, the relevance of fractionalization remains unclear, in particular, to the experimentally-observable quantities.

In this paper, we investigate the effect of fractionalization of quantum spins on the finite-$T$ properties of the Kitaev model on a honeycomb lattice by applying the unbiased quantum Monte Carlo (QMC) method. 
In this model, the two Majorana fermions, itinerant and localized, release their entropy successively at two well-separated $T$ scales.
We elucidate that each crossover has an impact on experimental observables: the high-$T$ one, driven by itinerant Majorana fermions, corresponds to the development of spin correlations between neighboring sites, while the low-$T$ one, originating from thermal fluctuations of localized Majorana fermions, is accompanied by a sizable change in the excitation spectrum of itinerant Majorana fermions. 
This leads to apparent $T$-linear behavior of the specific heat and coherent transport of Majorana fermions in the intermediate-$T$ state between the two crossovers, in contrast to the Dirac semimetallic behavior and $T^2$ specific heat anticipated in the low-$T$ limit.
Moreover, we show that the thermal excitation of fluxes tends to open a gap at finite $T$ near the gapless-gapped phase boundary.

The paper is structured as follows. In Sec.~\ref{sec:model}, we introduce the Kitaev model on a honeycomb lattice, and briefly review the ground-state properties.
In Sec.~\ref{sec:method}, we present the numerical method to analyze the finite-$T$ properties in the Kitaev model.
The definitions of physical quantities are also given in this section.
The numerical results are shown in Sec.~\ref{sec:results}.
We present the $T$ dependences of the specific heat and the entropy, and summarize the two crossovers in the phase diagram in Sec.~\ref{sec:specific-heat}.  
We discuss the origins of the crossovers by calculating the spin correlation in Sec.~\ref{sec:spin-correlation} and the flux density in Sec.~\ref{sec:flux-density}.
We also compute the density of states (DOS) of the itinerant Majorana fermions in Sec.~\ref{sec:density-states}.
We discuss the peculiar $T$ dependence of the specific heat in the intermediate-$T$ region in Sec.~\ref{sec:pecul-t-depend}.
In Sec.~\ref{sec:optic-cond-drude}, we evaluate the optical conductivity and the Drude weight of the itinerant Majorana fermions at finite $T$.
The effect of thermal fluctuations near the gapless-gapped boundary is discussed in Sec.~\ref{sec:gapl-gapp-bound}.
Finally, Sec.~\ref{sec:concluding-remarks} is devoted to the summary.

\section{Model}\label{sec:model}

The Kitaev model is composed of $S=1/2$ spins defined on a honeycomb lattice, whose Hamiltonian is given by~\cite{Kitaev06}
\begin{align}
 {\cal H}=-J_x\sum_{\means{jk}_x}\sigma_j^{x}\sigma_k^{x}-J_y\sum_{\means{jk}_y}\sigma_j^{y}\sigma_k^{y}-J_z\sum_{\means{jk}_z}\sigma_j^{z}\sigma_k^{z},
\label{eq:Horg}
\end{align}
where $\sigma_j^{l}$ is the $l(=x,y,z)$ component of the Pauli matrix representing an $S=1/2$ spin at site $j$. Corresponding to three inequivalent bonds on the honeycomb lattice, named $x$, $y$, and $z$ bonds, the sum over $\means{jk}_l$ is taken over the nearest neighbor (NN) sites on the $l$ bonds.

The ground state of the model in Eq.~(\ref{eq:Horg}) was exactly solved by introducing Majorana fermions~\cite{Kitaev06}. 
The ground state has gapped and gapless excitations depending on the exchange constants, $J_x$, $J_y$, and $J_z$~\cite{Kitaev06} (see the inset of Fig.~\ref{gap}).
The spin correlations are extremely short-ranged, i.e., nonzero only for the NN pairs for all the parameters, which indicates that both gapped and gapless ground states are QSLs~\cite{Baskaran2007,Schmidt2008,Vidal2008}.
The model does not exhibit any phase transition at finite $T$ although a three-dimensional variant does~\cite{Nasu2014a}.

Hereafter, we describe the anisotropy of the exchange constants by the parameter $\alpha$ as $J_x=J_y=\alpha/3$ and $J_z=1-2\alpha/3$ ($J_x+J_y+J_z=1$) as shown in the inset of Fig.~\ref{gap}.
Along this cut in the ground-state phase diagram, the gapped-gapless phase boundary in the ground state is located at $\alpha=3/4$.

\section{Method}\label{sec:method}

An exact solution for the ground state of the Kitaev model is formulated by the Jordan-Wigner transformation along the chains consisting of the $x$ and $y$ bonds~\cite{Chen2007,Feng2007,Chen2008}.
The fermions introduced by the transformation can be represented by two Majorana fermions $c_j$ and $\bar{c}_j$ at each site $j$.
Using these Majorana fermions, the Kitaev model is rewritten as 
\begin{align}
 {\cal H}=iJ_x \sum_{(jk)_x}c_j c_k
-iJ_y\sum_{(jk)_y}c_j c_k
-iJ_z\sum_{(jk)_z}\eta_r c_j c_k,\label{eq:1}
\end{align}
where the sum over $(jk)$ is taken for the NN sites with $j<k$.
The operator $\eta_r=i\bar{c}_j\bar{c}_k$ is defined on each $z$ bond ($r$ is the bond index).
This is regarded as a classical variable taking $\pm 1$ because of $[{\cal H},\eta_r]=0$ and $\eta_r^2=1$ for all $r$.

By using this Majorana representation, we carry out the QMC simulation at finite $T$. 
For the model in Eq.~(\ref{eq:1}), the partition function $Z$ is written in the form
\begin{align}
 Z={\rm Tr}_{\{\eta_r\}}{\rm Tr}_{\{c_i\}}e^{-\beta {\cal H}}=\sum_{\{\eta_r\}=\pm 1}e^{-\beta F_f(\{\eta_r\})},
\end{align}
where ${\rm Tr}_{\{\eta_r\}}$ and ${\rm Tr}_{\{c_i\}}$ are the traces for localized and itinerant Majorana fermions, respectively; $\beta=1/T$ is the inverse temperature (we set the Boltzmann constant $k_{\rm B}=1$).
Here, $F_f(\{\eta_r\})$ is the free energy of the Majorana fermion system for a fixed configuration $\{\eta_r\}$, which is easily calculated by the exact diagonalization. 
We perform the Markov chain Monte Carlo (MC) simulation for sampling the configurations of $\{\eta_r\}$ so as to reproduce the thermal distribution of $e^{-\beta F_f(\{\eta_r\})}$~\cite{Nasu2014a}.
In the present calculations, we performed the QMC simulation hybridized with the parallel tempering technique with 16 replicas~\cite{Hukushima1996}. 
We spent the 10,000 MC steps for thermalization and 40,000 MC steps for measurement in up to an $L=12$ cluster, which contains $N=2\times L^2=288$ sites.

In the QMC simulation, we calculate the specific heat as
\begin{align}
 C_v=\frac{1}{NT^2}
\left(
\means{E_f^2}-\means{E_f}^2-\mean{\frac{\partial E_f}{\partial \beta}}
\right),
\label{eq:Cv}
\end{align}
where $E_f(\{\eta_r\})$ is the energy of the itinerant Majorana fermion system for a given configuration $\{\eta_r\}$.
From the $T$ dependence of the specific heat, we obtain the entropy per site as
\begin{align}
 S=\ln 2 - \int_T^{T_m} dT' C_v/T',
\label{eq:entropy}
\end{align}
where $T_m$ is chosen to be $T_m=10\ (\gg J_x+J_y+J_z\equiv 1)$.

We also calculate the equal-time spin correlations. 
In the Kitaev model, as $\mean{\sigma_j^l \sigma_k^l}\neq 0$ for NN bonds~\cite{Baskaran2007}, we measure the NN spin correlations by 
\begin{align}
S^{ll}=\frac{2}{N} \sum_{\means{jk}_l}\means{\sigma_j^l\sigma_k^l},
\label{eq:spin_corr}
\end{align}
which are given by each term in Eq.~(\ref{eq:1}) in terms of the Majorana fermions.
In addition, we compute the thermal average of the flux density as
\begin{align}
W = \frac{2}{N} \sum_p \means{W_p},
\label{eq:W}
\end{align}
where $W_p$ is composed of $\eta_r$ included in the hexagonal plaquette $p$: $W_p=\prod_{r\in p}\eta_r$.

In addition to the above thermodynamic quantities, we calculate the dynamical quantities for itinerant Majorana fermions.
The DOS of the itinerant Majorana fermions with a given configuration of $\eta_r$ is defined by 
\begin{align}
 D(\omega,\{\eta_r\})=\sum_n\delta(\omega-\varepsilon_n(\{\eta_r\})), 
\end{align}
where $\varepsilon_n$ is the one-particle energy of the fermion $f_n$ which is introduced so as to diagonalize the Hamiltonian as
\begin{align}
 {\cal H}(\{\eta_r\})=\sum_n \varepsilon_n(\{\eta_r\})\left(f_n^\dagger f_n-\frac{1}{2}\right).
\end{align}
We calculate the thermal averages of the DOS, $D(\omega)=\mean{D(\omega, \{\eta_r\})}$ by the QMC simulation. 
Note that $D(\omega)$ do not contain the $T$ dependence of the Fermi distribution function: we take into account the effect of thermal fluctuations only on $\eta_r$.

Moreover, we compute the optical conductivity of itinerant Majorana fermions.
For this purpose, we introduce the Fourier transform of the Hamiltonian as
\begin{align}
 {\cal H}=\sum_{\bm{k}}\bm{c}_{\bm{k}}^\dagger H_{\bm{k}}\bm{c}_{\bm{k}}=\sum_{n:\varepsilon_{n\bm{k}}>0}\sum_{\bm{k}}\varepsilon_{n\bm{k}}\left(f_{n\bm{k}}^\dagger f_{n\bm{k}}-\frac{1}{2}\right),
\end{align}
where $\bm{c}_{\bm{k}}$ is a set of the Fourier transforms of $c_j$ and the $L\times L$ cluster is regarded as a unit cell.
The Bloch Hamiltonian $H_{\bm{k}}$ is diagonalized by introducing a set of fermions $f_{n\bm{k}}$ belonging to the $n$-th band with the energy $\varepsilon_{n\bm{k}}$.
Then, the conductivity tensor is calculated by
\begin{align}
 \sigma^{\mu\nu}(\omega)=
\frac{1}{L}\int_0^\infty dt e^{i(\omega+i\delta)t}\int_0^\beta d\lambda\means{J_\nu (-i\lambda)J_\mu(t)},
\label{eq:sigma_omega}
\end{align}
where $\delta$ is an infinitesimal positive number,
${\cal O}(t)=e^{i{\cal H}t}{\cal O}e^{-i{\cal H}t}$, and the current operator is defined as $J_{\mu}=\sum_{\bm{k}nn'}f_{\bm{k}n}^\dagger f_{\bm{k}n'}\bras{u_{\bm{k}n}}\partial H_{\bm{k}}/\partial k_{\mu}\kets{u_{\bm{k}n'}}$ with the eigenstate $\kets{u_{\bm{k}n}}$ of $H_{\bm{k}}$.
To extract the contribution to coherent transport,
we also obtain the Drude weight via the sum rule. Specifically, we compute the Drude weight along the $x$ direction by
\begin{align}
 D_x=\frac{1}{2L}\sum_{\means{ij}_x'}\means{J_x\sigma_i^x\sigma_j^x}
-\frac{1}{\pi}\int_0^{\infty}\sigma^{xx}(\omega)d\omega,\label{eq:2}
\end{align}
where the summation $\sum_{\means{ij}_x'}$ is taken only for the NN $x$ bonds on the boundary of a finite-size cluster which we regard as a unit cell in the calculations.

\section{Results}\label{sec:results}

\subsection{Specific heat and entropy}\label{sec:specific-heat}

\begin{figure*}[t]
\begin{center}
\includegraphics[width=2\columnwidth,clip]{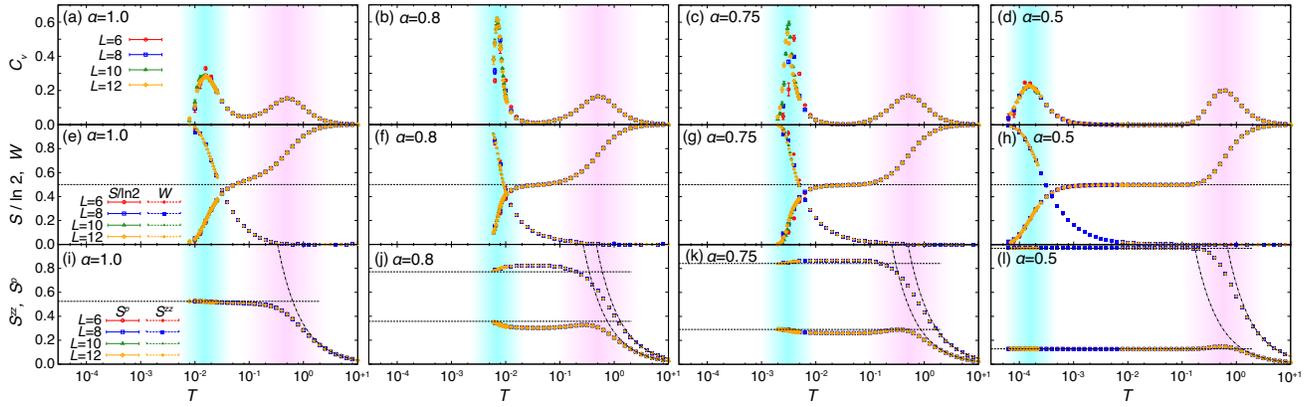}
\caption{(color online).
(a)-(d) $T$ dependences of the specific heat at (a) $\alpha=1.0$, (b) $\alpha=0.8$, (c) $\alpha=0.75$, and (d) $\alpha=0.5$ in the several clusters with $2\times L^2$ spins.
Here, we define the anisotropy parameter $\alpha$ by taking $J_x=J_y=\alpha/3$ and $J_z=1-2\alpha/3$.
(e)-(h) $T$ dependences of the entropy per site, $S$, and the thermal average of the density of the flux $W_p$, $W$.
(i)-(l) $T$ dependences of the equal-time spin correlations, $S^{ll}$; $S^p=(S^{xx}+S^{yy})/2$.
The horizontal dashed lines represent the values at $T=0$ which are calculated analytically~\cite{Baskaran2007}, and the dashed-dotted curves represent the high-$T$ Curie behaviors $S^{ll}\sim J_l/T$. 
}
\label{Cv}
\end{center}
\end{figure*}

Figures~\ref{Cv}(a)-\ref{Cv}(d) show the QMC data for the specific heat $C_v$ [Eq.~(\ref{eq:Cv})] as a function of $T$ for several values of the anisotropy parameter $\alpha$.
For all cases, the specific heat exhibits two peaks; both are almost system-size independent, indicating two crossovers.
We hereafter term the low- and high-$T$ crossover temperatures as $T_{\rm L}$ and $T_{\rm H}$, respectively.
Figures~\ref{Cv}(e)-\ref{Cv}(h) show the entropy per site obtained by Eq.~(\ref{eq:entropy}).
The entropy rapidly decreases with decreasing $T$ in the vicinity of $T_{\rm L}$ and $T_{\rm H}$ corresponding to the two peaks of the specific heat.
A half of the entropy is released successively in each crossover; consequently, the entropy becomes $\sim \frac{1}{2}\ln 2$ per site in the region between $T_{\rm L}$ and $T_{\rm H}$.
The plateau-like behavior of the entropy in this region becomes clearer for smaller $\alpha$, i.e., larger anisotropy of the exchange constants.

\begin{figure}[t]
\begin{center}
\includegraphics[width=\columnwidth,clip]{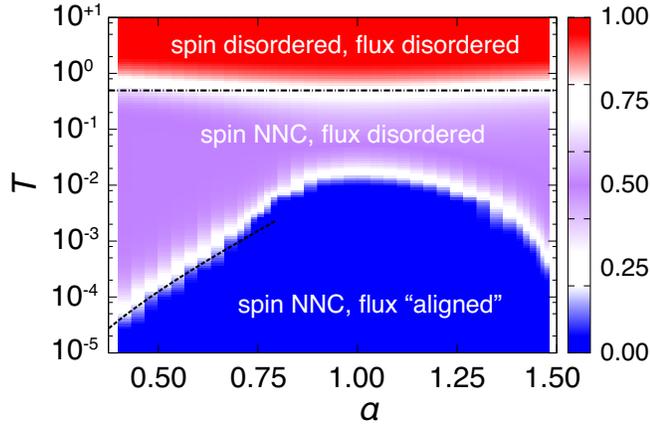}
\caption{(color online).
Contour map of the entropy per site, $S/\ln 2$, on a plane of $T$ and $\alpha$.
The dashed line represents crossover temperature obtained by the perturbation theory in the limit of $J_z\gg J_x,J_y$ ($\alpha\ll 1$).
The dashed-dotted line represents the crossover temperature obtained by assuming the constant DOS. 
NNC stands for a NN correlation.
}
\label{phase}
\end{center}
\end{figure}

Figure~\ref{phase} shows the contour map of the entropy on the $\alpha$-$T$ plane.
The high-$T$ crossover temperature $T_{\rm H}$ is almost independent of $\alpha$. 
The origin will be discussed in the next section \ref{sec:spin-correlation}.
On the other hand, the low-$T$ crossover temperature $T_{\rm L}$ strongly depends on $\alpha$; this will be discussed in Sec.~\ref{sec:flux-density}. 
There are three regions separated by the two crossovers $T_{\rm L}$ and $T_{\rm H}$.
In the following sections, we clarify the differences between these regions and their influences on the observable quantities.

\subsection{High-$T$ crossover: spin correlation}\label{sec:spin-correlation}

Let us first discuss what takes place in the high-$T$ crossover at $T_{\rm H}$.
The itinerant Majorana fermions $c_{j}$ form a band whose width is $W_B=2(J_x+J_y+J_z)-\Delta=2-\Delta$, where $\Delta$ is the excitation gap in the gapped phase.
Suppose that the system is in the gapless region and the DOS is constant $\sim 1/W_B$, the specific heat originating from the itinerant Majorana fermions takes maximum at $T\sim 0.511$. 
This value well coincides with $T_{\rm H}$ in a wide region of $\alpha$, even in the gapped region for $\alpha<0.75$, as shown by the dashed-dotted line in Fig.~\ref{phase}.
The result clearly indicates that the high-$T$ crossover originates in the itinerant Majorana fermions.

We find that the crossover at $T_{\rm H}$ is closely related with the development of NN spin correlations given in Eq.~(\ref{eq:spin_corr}), which is observable in experiments.
The $T$ dependences of $S^{ll}$ are presented in Figs.~\ref{Cv}(i)-\ref{Cv}(l) for the same set of $\alpha$ as in Figs.~\ref{Cv}(a)-\ref{Cv}(h).
In the high-$T$ limit, $S_{jk}^{ll}$ is given by the high-$T$ expansion as ${\rm Tr}[\sigma_j^l \sigma_k^l e^{-\beta {\cal H}}]/{\rm Tr}e^{-\beta {\cal H}}\sim -\beta {\rm Tr}[\sigma_j^l\sigma_k^l {\cal H}]=\beta J_l$.
Our QMC data obey this Curie behavior, indicated by the dashed-dotted curves in Figs.~\ref{Cv}(i)-\ref{Cv}(l).
In the crossover region near $T_{\rm H}$, however, the spin correlations show deviations from the Curie behavior, and quickly saturate to the values that are analytically obtained for the ground state (horizontal dashed lines in the figures)~\cite{Baskaran2007}.
Hence, the high-$T$ crossover by the itinerant Majorana fermions corresponds to physically important behavior in this quantum spin system: the growth of the NN spin correlations.
We note that the spin correlations also show slight changes in the low-$T$ crossover at $T_{\rm L}$. This behavior is discussed in Sec.~\ref{sec:gapl-gapp-bound}.

\subsection{Low-$T$ crossover: flux density}\label{sec:flux-density}

Next, we discuss what occurs in the low-$T$ crossover. 
The entropy release near $T_{\rm L}$ originates from the localized Majorana fermions $\bar{c}_j$ or $\eta_r$.
This is confirmed by calculating the $T$ dependence of the flux density, $W$, in Eq.~(\ref{eq:W}), as shown in Figs.~\ref{Cv}(e)-\ref{Cv}(h). 
The results show that $W$ rapidly decreases from $1$ with increasing $T$ in the vicinity of $T_{\rm L}$.
Hence, the crossover at $T_{\rm L}$ is due to the thermal fluctuation of fluxes. 

This is further confirmed by considering the toric code limit corresponding to $J_x,J_y\ll J_z$ ($\alpha \ll 1$). 
In this limit, the Kitaev model is reduced to the effective model ${\cal H}_{\rm eff}=-J_{\rm eff}\sum_p W_p$, where $J_{\rm eff}=J_x^2J_y^2/(16J_z^3)$~\cite{Kitaev06}. 
Since this effective model describes free Ising spins in the magnetic field $J_{\rm eff}$, the specific heat is of Schottky-type, which takes a maximum at $\tilde{T}_{\rm L}/J_{\rm eff}\sim 0.833$.
The asymptotic behavior is shown by the dashed line in Fig.~\ref{phase}.
The agreement between this line and $T_{\rm L}$ further supports that the low-$T$ crossover is caused by the thermal disturbance of  fluxes.

We also note that the agreement of the asymptotic behavior is consistent with the absence of phase transition in this two-dimensional system.
This is in contrast to the three-dimensional case; there are local constraints for $W_p$ in the Kitaev model defined on a three-dimensional hyperhoneycomb lattice, leading to a finite-$T$ phase transition~\cite{Nasu2014a}. 
On the other hand, there is no constraint for $W_p$ in the two-dimensional case, which results in the absence of the phase transition for $T>0$.

To summarize the above results, the three regions in the phase diagram depicted in Fig.~\ref{phase} are characterized as follows.
The high-$T$ region for $T \gtrsim T_{\rm H}$ is a conventional paramagnetic state, where the NN spin correlations obey the Curie behavior. 
On the other hand, in the low-$T$ region for $T<T_{\rm L}$, the NN spin correlations are saturated to the $T=0$ values, and furthermore, the fluxes are also aligned. 
Thus, the system below $T_{\rm L}$ behaves similar to the ground state QSL. 
In the region for $T_{\rm L} \lesssim T \lesssim T_{\rm H}$, a peculiar intermediate state appears: the NN spin correlations are well developed, whereas the fluxes are thermally disordered. 
In the following sections, we discuss the nature of this intermediate state.

\subsection{Density of states of itinerant Majorana fermions}\label{sec:density-states}

\begin{figure*}[t]
\begin{center}
\includegraphics[width=2\columnwidth,clip]{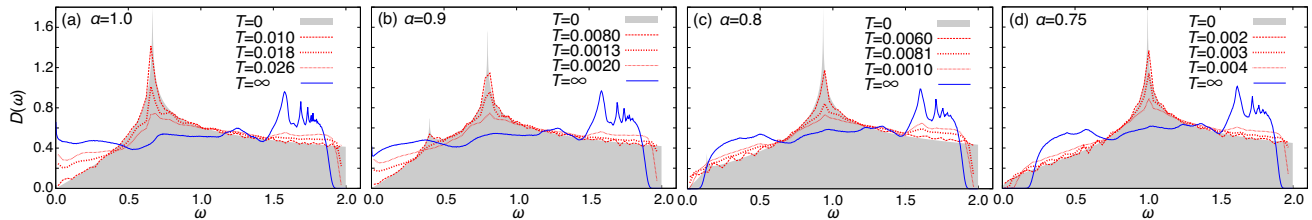}
\caption{(color online).
The DOS of Majorana fermions at (a) $\alpha=1.0$, (b) $\alpha=0.9$, (c) $\alpha=0.8$, and (d) $\alpha=0.75$. 
Except the results at $T=0$ and $T=\infty$, the DOS are calculated by QMC for the $10\times 10$ superlattice of the $L=12$ cluster.
}
\label{dos}
\end{center}
\end{figure*}

Since the $Z_2$ variables $\eta_r$ couple with the itinerant Majorana fermions, we expect that the enhanced fluctuations of fluxes above $T_{\rm L}$ affect the nature of itinerant Majorana fermions considerably.
In order to elucidate such behavior, we calculate the DOS of itinerant Majorana fermions. 
The calculations were done for the $10\times 10$ supercell, where the $L=12$ cluster obtained by the MC simulation is regarded as a unit cell. 
The calculations at $T=0$ ($T=\infty$) are performed for a $L=6,000$ ($L=60$) cluster.
In the calculation at $T=\infty$, we take a simple average over 10,000 random configurations of $\{\eta_r\}$.

Figure~\ref{dos}(a) shows the result for the isotropic case $\alpha=1.0$ ($J_x=J_y=J_z$). 
The QMC data are shown near $T_{\rm L}$, together with the results at $T=0$ and $T=\infty$. 
In this gapless case, at $T=0$, the DOS shows semimetallic behavior $D(\omega)=\mean{D(\omega, \{\eta_r\})} \propto \omega$ for small $\omega$, reflecting the Dirac dispersion. 
While increasing $T$ above $T_{\rm L}$, however, the semimetallic dip of DOS is filled rapidly, leading to ``metallic'' behavior, $D(\omega=0) \neq 0$. 
The result clearly indicates that the thermal fluctuations of fluxes near $T_{\rm L}$ significantly affect the low-energy spectrum of itinerant Majorana fermions.

\subsection{Peculiar $T$ dependence of the specific heat}\label{sec:pecul-t-depend}

\begin{figure}[t]
\begin{center}
\includegraphics[width=\columnwidth,clip]{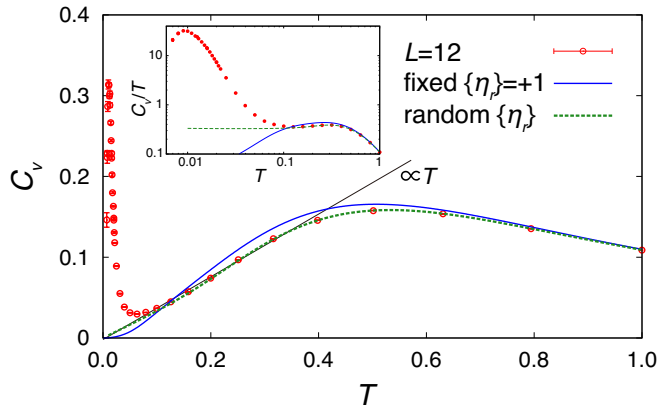}
\caption{(color online).
$T$ dependence of the specific heat $C_v$ at $\alpha=1.2$ in the $L=12$ cluster.
 For comparison, the results calculated by fixing all $\eta_r$ to $+1$ and by assuming random $\{\eta_r\}$ are shown 
 by the solid and dashed curves, respectively.
 The log plot of $C_v/T$ is also shown in the inset.
}
\label{Cv1.2}
\end{center}
\end{figure}

The significant change in the DOS for $T>T_{\rm L}$ affects the $T$ dependence of the specific heat $C_v$.
In the gapless QSL region, the low-$T$ specific heat is expected to be proportional to $T^2$ because of the Dirac semimetallic dispersion for aligned fluxes.
However, $C_v$ calculated by assuming all $\eta_r=+1$ largely deviates from our QMC data in the calculated $T$ range, as shown for $\alpha=1.2$ in Fig.~\ref{Cv1.2}.
This indicates that the asymptotic $T^2$ behavior will be limited only in the extremely low-$T$ region, much lower than $T_{\rm L}$. 

Instead, in a wide range of $T_{\rm L} \lesssim T \lesssim T_{\rm H}$, we find that $C_v$ well scales to $\propto T$, which originates from the ``metallic'' DOS caused by thermally fluctuating fluxes above $T_{\rm L}$. 
Indeed, the overall behavior including $T \gtrsim T_{\rm H}$ is well explained by the result for completely random $\{\eta_r \}$, as shown in Fig.~\ref{Cv1.2}.
Thus, as a consequence of the thermal fractionalization of quantum spins, we find the apparent $T$-linear behavior, not $T^2$, in the intermediate-$T$ region where the NN spin correlations are well developed.

\subsection{Optical conductivity and Drude weight}\label{sec:optic-cond-drude}

\begin{figure}[t]
\begin{center}
\includegraphics[width=\columnwidth,clip]{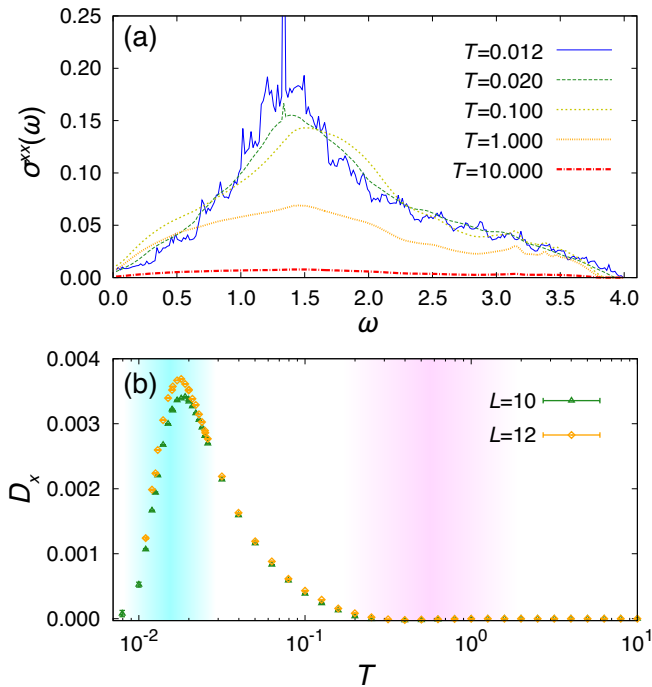}
 \caption{(color online).
 (a) The optical conductivity at $\alpha=1.0$ on the $L=12$ cluster at several $T$.
 (b) $T$ dependence of the Drude weight of itinerant Majorana fermions at $\alpha=1.0$.
}
\label{cond}
\end{center}
\end{figure}

The significant change in the DOS of the itinerant Majorana fermions will affect transport properties as well.
We here show it by computing the optical conductivity of itinerant Majorana fermions [Eq.~(\ref{eq:sigma_omega})].
Here, we consider the longitudinal component along the $x$ direction ($\mu=\nu=x$). 
The calculations were done for the $1\times 1$ supercell of the $L=12$ cluster.

Figure~\ref{cond}(a) shows the results of $\sigma^{xx}(\omega)$ at several $T$ for $\alpha=1.0$.
The incoherent component at finite $\omega$ increases with decreasing $T$ below $T_{\rm H}$.
To extract the contribution to coherent transport,
we calculate the Drude weight $D_x$ of the itinerant Majorana fermions by using the sum rule in Eq.~(\ref{eq:2}).
Figure~\ref{cond}(b) shows the $T$ dependence of $D_x$.
While decreasing $T$, the Drude weight gradually increases below $T_{\rm H}$, and sharply decreases to zero below $T_{\rm L}$ after showing a peak near $T_{\rm L}$.
The result suggests that the transport quantities, such as the thermal conductivity, have sizable values between the two crossovers.

\subsection{Gapless-gapped phase boundary}
\label{sec:gapl-gapp-bound}

\begin{figure}[b]
\begin{center}
\includegraphics[width=\columnwidth,clip]{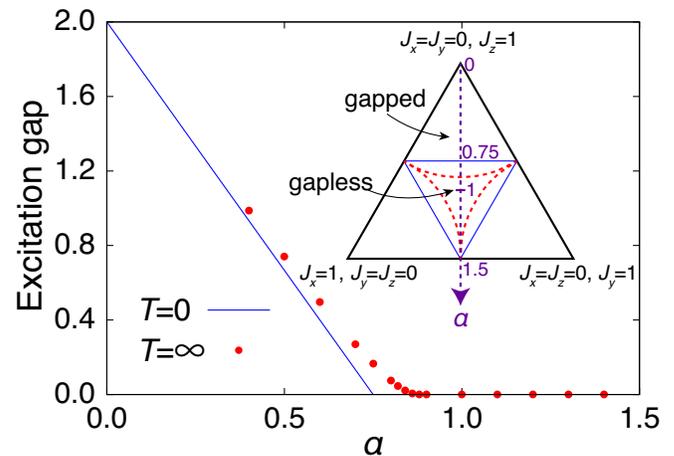}
\caption{(color online).
The excitation gap for the Majorana fermions at $T=0$ (blue solid line) and $T=\infty$ (red symbols) as a function of $\alpha$.
The inset indicates the gapped-gapless boundaries on the plane of $J_x+J_y+J_z=1$. 
The blue solid lines represent the phase boundaries in the ground state, while the red dashed lines represent the boundaries obtained from the DOS at $T=\infty$.
See the text for details.
}
\label{gap}
\end{center}
\end{figure}

Finally, we discuss the effect of thermal fluctuations near the phase boundary between the gapless and gapped phases. 
Whereas the coherent transport and the $T$-linear behavior are observed widely in the region where the ground state is gapless, they are disturbed in the vicinity of the phase boundary at $\alpha=0.75$.
In this region, thermal fluctuations in the $Z_2$ variables $\{\eta_r\}$ bring about different behavior in the low-energy part of $D(\omega)=\mean{D(\omega, \{\eta_r\})}$.
Figures~\ref{dos}(b), \ref{dos}(c), and \ref{dos}(d) show the DOS of the itinerant Majorana fermions at $\alpha=0.9$, $\alpha=0.8$, and $\alpha=0.75$, respectively.
At $\alpha=0.8$ and $\alpha=0.75$, the system develops an energy gap with increasing $T$ in the vicinity of $T_{\rm L}$, in sharp contrast to the gap filling at $\alpha=1.0$ and $\alpha=0.9$.
The results indicate that there is an intermediate region where the thermal fluctuation of $\eta_r$ gaps out the low-energy excitation of itinerant Majorana fermions.

The intermediate region is identified by calculating the magnitudes of the gaps at $T=0$ and $T=\infty$, as presented in Fig.~\ref{gap}.
The schematic phase diagram determined by the DOS at $T=\infty$ is presented in the inset.
Remarkably, the gapped-gapless boundary is similar to that in the dynamical phase diagram~\cite{Knolle2014}, suggesting a relation between thermal and quantum fluctuations.
We also note that the boundary is similar to the result for the full flux state~\cite{Lahtinen2012}.

The modification of the boundary at finite $T$ implies that effective exchange couplings are renormalized in an anisotropic way by the thermal fluctuation of the $Z_2$ variables $\{\eta_r\}$. 
Indeed, the anisotropy of spin correlations is slightly enhanced near $T_{\rm L}$ while increasing $T$, as shown in Figs.~\ref{Cv}(j) and \ref{Cv}(k).
Thus, the slight change in the spin anisotropy is interpreted as a consequence of the change of the excitation gap in the itinerant Majorana fermions fractionalized from the quantum spins.

\section{Summary}\label{sec:concluding-remarks}

In summary, we have investigated the thermal fractionalization of quantum spins into Majorana fermions in the Kitaev model by using the QMC simulation.
We clarified that the fractionalization appears as two crossovers, both of which are physically observable in the thermodynamics. 
The higher-$T$ crossover is identified by the development of NN spin correlations, which will be observed in, e.g., neutron scattering experiments.
In between the crossovers, the Drude weight of itinerant Majorana fermions takes a sizable value, which might be observed by the thermal conductivity.
Meanwhile, the low-$T$ crossover leads to a peculiar $T$-linear behavior in the specific heat above the crossover temperature.
We also showed that the thermal fractionalization affects the gapped-gapless phase boundary by renormalizing the spin anisotropy.
The present results complete how the fractionalization of quantum spins into Majorana fermions occurs while changing temperature in the ideal Kitaev model. 
This provides a useful reference to the experimental exploration of QSLs in, e.g., iridium oxides~\cite{Singh2010,Singh2012,Comin2012,Choi2012,Ohgushi2013} and ruthenium compounds~\cite{Plumb2014,Kubota2015,Sears2014,Majumder2014}, where the dominant interaction is expected to be of Kitaev type.

\begin{acknowledgments}
This work is supported by Grant-in-Aid for Scientific Research, the Strategic Programs for Innovative Research (SPIRE), MEXT, and the Computational Materials Science Initiative (CMSI), Japan.
Parts of the numerical calculations are performed in the supercomputing systems in ISSP, the University of Tokyo.
\end{acknowledgments}


\begin{thebibliography}{99} 

\bibitem{Tomonaga1950}
S. Tomonaga,
\journal{\PTP}{5}{544}{1950}.

\bibitem{Wen1991}
X. G. Wen,
\journal{\prb}{44}{2664}{1991}.

\bibitem{Nussinov2007}
Z. Nussinov, C. D. Batista, B. Normand, and S. A. Trugman,
\journal{\PRB}{75}{094411}{2007}.

\bibitem{Normand2014}
B. Normand and Z. Nussinov,
\journal{\PRL}{112}{207202}{2014}.

        
\bibitem{Castelnovo2008}
C. Castelnovo, R. Moessner, and S. L. Sondhi,
\journal{Nature}{451}{42}{2008}.


\bibitem{Emery1992}
V. J. Emery and S. Kivelson
\journal{\PRB}{46}{10812}{1992}.




\bibitem{Kitaev06}
A.~Kitaev,
\journal{Ann. Phys.}{321}{2}{2006}.


\bibitem{Jackeli2009}
G. Jackeli and G. Khaliullin,
\journal{\PRL}{102}{017205}{2009}. 




\bibitem{Chen2007}
H.-D. Chen and J. Hu,
\journal{\PRB}{76}{193101}{2007}.

\bibitem{Feng2007}
X.-Y. Feng, G.-M. Zhang, and T. Xiang, 
\journal{\PRL}{98}{087204}{2007}.

\bibitem{Chen2008}
H.-D. Chen, and Z. Nussinov, 
\journal{J. Phys. A Math. Theor.}{41}{075001}{2008}.



\bibitem{Chaloupka2010}
J. Chaloupka, G. Jackeli, and G. Khaliullin,
\journal{\PRL}{105}{027204}{2010}.

\bibitem{Reuther2011}
J. Reuther, R. Thomale, and S. Trebst,
\journal{\PRB}{84}{100406}{2011}.

\bibitem{Chaloupka2013}
J. Chaloupka, G. Jackeli, and G. Khaliullin,
\journal{\PRL}{110}{097204}{2013}.



\bibitem{Nasu2014}
J. Nasu, T. Kaji, K. Matsuura, M. Udagawa, and Y. Motome, 
\journal{\PRB}{89}{115125}{2014}.

\bibitem{Nasu2014a}
J. Nasu, M. Udagawa, and Y. Motome, 
\journal{\PRL}{113}{197205}{2014}.

\bibitem{Nasu2014b}
J. Nasu, M. Udagawa, and Y. Motome, 
\journal{J. Phys.: Conf. Ser.}{592}{012115}{2015}.



\bibitem{Baskaran2007}
G.~Baskaran, S.~Mandal, and R.~Shankar,
\journal{\PRL}{98}{247201}{2007}.

 \bibitem{Schmidt2008}
K. P. Schmidt, S. Dusuel, and J. Vidal,
\journal{\PRL}{100}{057208}{2008}.

 \bibitem{Vidal2008}
J. Vidal, K. P. Schmidt, and S. Dusuel,
\journal{\PRB}{78}{245121}{2008}.



\bibitem{Hukushima1996}
K. Hukushima and K. Nemoto,
\journal{\JPSJ}{65}{1604}{1996}.

\bibitem{Knolle2014}
J. Knolle, D. L. Kovrizhin, J. T. Chalker, and R. Moessner,
\journal{\PRL}{112}{207203}{2014}.


\bibitem{Lahtinen2012}
V. Lahtinen, G. Kells, A Carollo, T. Stitt, J. Vala, J. K. Pachos,
\journal{Ann. Phys.}{323}{2286}{2008}.


\bibitem{Singh2010}
Y. Singh and P. Gegenwart,
\journal{\PRB}{82}{064412}{2010}.

\bibitem{Singh2012}
Y. Singh, S. Manni, J. Reuther, T. Berlijn, R. Thomale, W. Ku, S. Trebst, and P. Gegenwart,
\journal{\PRL}{108}{127203}{2012}.

\bibitem{Choi2012}
S. K. Choi, R. Coldea, A. N. Kolmogorov, T. Lancaster, I. I. Mazin, S. J. Blundell, P. G. Radaelli, Y. Singh, P. Gegenwart, K. R. Choi, S.-W. Cheong, P. J. Baker, C. Stock, and J. Taylor,
\journal{\PRL}{108}{127204}{2012}.


\bibitem{Comin2012}
R. Comin, G. Levy, B. Ludbrook, Z.-H. Zhu, C. N. Veenstra, J. A. Rosen, Y. Singh, P. Gegenwart, D. Stricker, J. N. Hancock, D. van der Marel, I. S. Elfimov, and A. Damascelli,
\journal{\PRL}{109}{266406}{2012}.


\bibitem{Ohgushi2013}
K. Ohgushi, J. I. Yamaura, H. Ohsumi, K. Sugimoto, S. Takeshita, A. Tokuda, H. Takagi, M. Takata, and T. H. Arima,
\journal{\PRL}{110}{217212}{2013}.


\bibitem{Plumb2014}
K. W. Plumb, J. P. Clancy, L. J. Sandilands, V. V. Shankar, Y. F. Hu, K. S. Burch, H. Y. Kee, and Y. J. Kim,
\journal{\PRB}{90}{041112}{2014}.

\bibitem{Kubota2015}
Y. Kubota, H. Tanaka, T. Ono, Y. Narumi, and K. Kindo,
\journal{\PRB}{91}{094422}{2015}.


\bibitem{Sears2014}
J. A. Sears, M. Songvilay, K. W. Plumb, J. P. Clancy, Y. Qiu, and Y. Kim,
arXiv:1411.4610.


\bibitem{Majumder2014}
M. Majumder, M. Schmidt, H. Rosner, A. A. Tsirlin, H. Yasuoka, and M. Baenitz,
arXiv:1411.6515.



\end{thebibliography}


\end{document}